\documentclass[t12pt,prd,reprint,preprintnumbers]{revtex4}
\usepackage{graphicx}
\usepackage{amsmath}
\usepackage{amssymb}

\makeatletter
\newcommand*{\rom}[1]{\expandafter\@slowromancap\romannumeral #1@}
\makeatother
\newcommand{\deltl}{\delta\mathcal{L}}
\newcommand{\parl}{\partial\mathcal{L}}
\newcommand{\galph}{g^{\alpha\beta}}

\begin{document}
\title{Fake Conformal Symmetry in Conformal Cosmological Models}
\author{R. Jackiw}
\affiliation{Department of Physics\\ Massachusetts Institute of Technology\\ Cambridge, Massachusetts 02139}
\author{So-Young Pi}
\affiliation{Department of Physics\\ 
Boston University, Boston, Massachusetts, 02215}


\begin{abstract}
We examine the local conformal invariance (Weyl invariance) in tensor-scalar theories used in recently proposed conformal cosmological models. We show that the Noether currents associated with Weyl invariance in these theories vanish. We assert that the corresponding Weyl symmetry does not have any dynamical role.
\end{abstract}

\maketitle

\section{Introduction}
Field theoretic models that possess Weyl invariance (local conformal invariance) are the focus of present day attention. Recently some cosmologists suggested in a series of papers that Weyl invariant dynamics can assist in unraveling various cosmological issues \cite{Kallosh:2013hoa, Bars:2013yba}. The simplest studied examples involve one or two scalar fields conformally coupled to the Ricci scalar $R$. The action for one-scalar-field model is given by
\begin{equation}
\begin{aligned}
I_1 &= - \int d^4 x \mathcal{L}_1\\[.5ex]
\mathcal{L}_1 &= \sqrt{-g} \left(\frac{1}{12}\ R \varphi^2 + \frac{1}{2}\ g^{\alpha\beta} \partial_\alpha \varphi\, \partial_\beta\, \varphi - \frac{1}{4}\ \lambda\varphi^{4}\right) .
\label{jackso1}
\end{aligned}
\end{equation}
 Henceforth the self-coupling is omitted, $\lambda =0$, since it has no bearing on our investigation. A two-field generalization of $I_1$ based on two scalar fields, $\varphi$ and $\psi$, conformally coupled to $R$ in an SO(1, 1) invariant manner has been posited in Ref. \cite{Bars:2013ybb}:
 \begin{equation}
\begin{aligned}
I_2 &= - \int d^4 x \mathcal{L}_2\\[.5ex]
\mathcal{L}_2 &= \sqrt{-g} \left\{\frac{1}{12}\ R (\varphi^2 - \psi^2) + \frac{1}{2}\ g^{\alpha\beta} \, (\partial_\alpha \varphi\, \partial_\beta\, \varphi - \partial_\alpha \, \psi \, \partial_\beta\, \psi)\right\} 
\label{jackso1-2}
\end{aligned}
\end{equation}
The action is invariant under the local Weyl transformation of the fields:
\begin{equation}
\begin{aligned}
 & g^{\alpha\beta} (x) \to e^{2\theta (x)} g^{\alpha\beta} (x),     \quad \delta g^{\alpha\beta} (x) = 2\theta (x) g^{\alpha\beta} (x) \\[1ex]
&    \varphi (x) \to e^{\theta (x)} \varphi (x),  \hspace{1.8em}  \delta \varphi (x) = \theta (x) \varphi (x)   \\[1ex]
&     \psi (x) \to e^{\theta (x)}  \psi (x), \hspace{1.8em}  \delta \psi(x) = \theta (x) \psi (x)
\label{jackso2}
     \end{aligned}
\end{equation}
In view of the claimed importance of the above Weyl invariance,  in this paper we examine in  detail the dynamical role of Weyl symmetry by determining its Noether symmetry current. Although Weyl symmetry has a long history \cite{PAM1973}, the associated current has not been previously studied (to our knowledge). We find that the current vanishes. 

In Sec.\! \rom{2} we present two calculations of Noether current using the one-scalar-field model \eqref{jackso1}. The two-scalar-field model \eqref{jackso1-2} behaves similarly, except that there is additional SO(1,1) symmetry, which we shall also discuss. We interpret our result in Sec. \rom{3}.
\eject
\section{Calculations of Noether currents}

\begin{description}
  \item[A.] Covariant calculation of Weyl current
\end{description}

We describe our calculation of the Weyl symmetry current associated with model \eqref{jackso1}, using the Noether procedure. (The subscript 1 in $\mathcal{L}$ is omitted in this section.) In its familiar form, Noether's first theorem deals with Lagrangians that depend at most on single derivatives of the dynamical fields. In our application the Lagrangian involves double derivatives (of $g^{\alpha\beta}$ in $R$). Thus some modification is needed \cite{Dbak2012ph}.

Without using the equations of motion the variation $\delta\mathcal{L}$ is 
\begin{subequations}
\begin{equation}
\deltl = \frac{\parl}{\partial\varphi}\ \delta \varphi + \frac{\parl}{\partial(\partial_\mu \varphi)}\ \partial_\mu \delta \varphi + \frac{\parl}{\partial\galph}\ \delta \galph +\frac{\parl}{\partial(\partial_\mu \galph)}\ \partial_\mu\, \delta \galph + \frac{\parl}{\partial(\partial_\mu\partial_\nu \galph)}\ \partial_\mu\partial_\nu \delta \galph .
\label{jackso3a}
\end{equation}
For the Lagrangian \eqref{jackso1} and the transformations \eqref{jackso2}, $\deltl$ in \eqref{jackso3a} is found to be
\begin{eqnarray}
\deltl & =& \partial_\mu X^\mu \label{jackso3b-1}\\[.5ex]
X^\mu & =& \frac{1}{2}\ \sqrt{- g}\, \varphi^2 g^{\mu\nu} \partial_\nu \theta .
\label{jackso3b}
\end{eqnarray}
\end{subequations}
where the evaluation of $X^\mu$ follows from the well-known scaling property of $\sqrt{-g} R$. Equation \eqref{jackso3b-1} is a consequence of the action being invariant against the transformations \eqref{jackso2}.

Next the Euler-Lagrange equations of motion
\begin{equation}
\begin{aligned}
\frac{\parl}{\partial\varphi} \ & =\ \partial_\mu \ \frac{\parl}{\partial(\partial_\mu \varphi)}, \\[1ex]
\frac{\parl}{\partial\, \galph}  \ & =\ \partial_\mu \ \frac{\parl}{\partial(\partial_\mu\, \galph)} -\partial_\mu \partial_\nu \ \frac{\parl}{\partial(\partial_\mu\, \partial_\nu\, \galph)} .
\label{jackso4}
\end{aligned}
\end{equation}
are used to eliminate $\frac{\parl}{\partial\varphi}$ and $\frac{\parl}{\partial\, \galph}$ from \eqref{jackso3a} thereby arriving at an alternate divergence formula for $\deltl$.
\begin{subequations}
\begin{alignat}{2}
\deltl  & = \partial_\mu\, K^\mu \label{jackso5a}\\[1ex]
K^\mu  & =  \frac{\parl}{\partial(\partial_\mu \varphi)}\ \delta\varphi + \frac{\parl}{\partial(\partial_\mu\, \galph)}\ \delta \galph + \frac{\parl}{\partial(\partial_\mu\, \partial_\nu\, \galph)}\ \partial_\nu \, \delta \galph - \partial_\nu\frac{\parl}{\partial(\partial_\mu \partial_\nu\, \galph)}\ \delta \galph \label{jackso5b}
\end{alignat}
\end{subequations}
Note that using the equations of motion always gives (\ref{jackso5a}, \ref{jackso5b}) regardless whether one is dealing with a symmetry transformation or not.

Equating the two formulas for $\deltl$ shows that the symmetry current
\begin{subequations}
\begin{equation}
J^\mu = K^\mu - X^\mu
\label{jackso6a}
\end{equation}
is conserved,
\begin{equation}
\partial_\mu\, J^\mu = 0.
\label{jackso6b}
\end{equation}
\end{subequations} 

A lengthy and tedious evaluation of \eqref{jackso5b} yields the simple result
\begin{subequations}
\begin{equation}
K^\mu = X^\mu \label{jackso7a}
\end{equation}
\vspace{-2ex}
and  the current vanishes:
\vspace{-2ex}
\begin{equation}
J^\mu = 0. \label{jackso7b}
\end{equation}
\end{subequations}

%

\begin{description}
  \item[B.] Noncovariant calculation of Weyl current
\end{description}
The occurrence of second-order derivatives of $\galph$ in $R$ is responsible for much of the tedium in our calculation. Therefore, it is useful to give a formulation in which double derivatives are absent. This is possible owing to the following identity satisfied by $R$, in which double derivatives are isolated:
\begin{subequations}
\begin{eqnarray}
\sqrt{-g}\, R &=& A+B = A + \partial_\alpha\, C^\alpha \label{jackso9a}\\[1ex]
	       A &=& \sqrt{-g}\, g^{\sigma\rho} \left(\Gamma^\lambda_{\sigma\kappa} \Gamma^\kappa_{\rho\lambda} - \Gamma^\kappa_{\sigma\rho} \, \Gamma^\lambda_{\kappa\lambda}\right)\label{jackso9b}\\[1ex]
	       C^\alpha &=& \sqrt{-g}\, \left(g^{\sigma\rho}\, \Gamma^\alpha_{\sigma\rho} - g^{\sigma \alpha}\, \Gamma^\lambda_{\sigma\lambda}\right)\label{jackso9c}
\end{eqnarray}
\end{subequations}
Here $A$ is free of double derivatives; they are contained in $B$, which is given by the divergence of $C^\alpha$. The latter depends solely on first derivatives of $\galph$. Thus
\begin{subequations}
\begin{equation}
\mathcal{L} = \frac{1}{12}\ \partial_{\alpha} (C^\alpha \varphi^2) + \mathcal{L}^\prime ,
\label{jackso10a}
\end{equation}
where
\begin{equation}
\mathcal{L}^\prime = \frac{1}{12}\  A\varphi^2 - \frac{1}{12}\ C^\alpha \partial_\alpha \, \varphi^2 + \sqrt{-g}\, \left(\frac{1}{2}\ \galph\, \partial_\alpha\, \varphi\, \partial_{\beta} \varphi \right) .
\label{jackso10b}
\end{equation}
\end{subequations}
Total derivative terms in Lagrangians have no effect on dynamics in the bulk. Therefore the argument can be based on $ \mathcal{L}^\prime$, which is free of second derivatives. 

Before proceeding, we first observe that the variation of $\deltl$ given in \eqref{jackso3b}, comes entirely from the total derivative term in \eqref{jackso10a}.
\begin{equation}
\deltl = \partial_{\mu} \left(\frac{1}{2}\ \sqrt{-g}\, \varphi^2\, g^{\mu\nu} \partial_\nu \theta\right) = \frac{1}{12}\ \partial_\mu\ [\delta(C^\mu\, \varphi^2)]
\label{jackso11}
\end{equation}
Hence another advantage of working with $\mathcal{L}^\prime$ is that it is invariant:
\begin{equation}
\deltl^\prime = 0 \quad \Rightarrow \quad X^{\prime\mu} = 0.
\label{jackso12}
\end{equation}
We again use Noether theorem with $\mathcal{L}^\prime$. The argument proceeds as before. One finds
\begin{equation}
K^{\prime\mu} = \left( \frac{\partial\mathcal{L}^\prime}{\partial(\partial_\mu \varphi)}\ \varphi + 2\ \frac{\partial\mathcal{L}^\prime}{\partial(\partial_{\mu}\, \galph)}\ \galph\right) \ \theta = 0.
\label{jackso13}
\end{equation}
Not only does the symmetry current vanish, but additionally $K^{\prime\mu}$ and $X^{\prime\mu}$ vanish separately.  

\begin{description}
  \item[C.] SO (1,1) symmetry
\end{description}
As is well known model (2) possesses global SO(1,1) symmetry which acts on the two scalar fields $\varphi$ and $\psi$:
\begin{eqnarray}
\delta \varphi (x) &=& \epsilon\, \psi (x)\nonumber\\
\delta \psi(x) &=& \epsilon\, \varphi (x)
\label{jackso14}
\end{eqnarray}
The current is readily determined by Noether's method:
\begin{equation}
J^\mu_{SO(1,1)} (x) = \epsilon \sqrt{-g}\,g^{\mu\nu} \, \left(\psi(x) \, \partial_\nu \varphi (x) - \varphi (x) \partial_\nu \psi (x)\right)
\label{jackso15}
\end{equation}
This is conserved by virtue of scalar field equations of motion.

\section{Interpretation of the result and discussion}

The Noether procedures always leave current formulas ambiguous up to identically conserved superpotentials. This is because one is extracting an expression from its divergence: $\partial_\mu (K^\mu - X^\mu) =0$ suggests that the conserved current is $J^\mu = K^\mu -X^\mu$.  In spite of the ambiguity due to the possible presence of superpotentials, the fact that  $K^{\prime \mu}$ and $X^{\prime \mu}$  vanish individually (in the noncovariant calculation) is strong evidence that current vanishes.

Moreover, an extension of Noether's theorem, called the ``second theorem," establishes that the current associated with a local symmetry is always a superpotential \cite{Dbak2012ph, Noe1918}. We applied Noether's second theorem to the model \eqref{jackso1} and regained our previous result: vanishing current.

The fact that the Weyl current vanishes cannot be attributed to the locality of the symmetry transformation parameter $\theta(x)$. An instructive example is electrodynamics, where $\delta A_{\mu} =\partial_{\mu} \theta$ and $\delta\Psi =- i\, \theta\, \Psi$ for a charged field $\Psi$. The current is nonvanishing and is identically conserved,  i.e. it is a superpotential.
\begin{equation}
J^\mu = \partial_\nu\, (F^{\mu\nu}\, \theta)
\label{jackso16}
\end{equation}
(This is the Noether current for gauge symmetry, not the source current $J^\mu_{EM}$  that appears in the Maxwell equations.) While the gauge dependence in \eqref{jackso16} with inhomogeneous $\theta$ may make $J^\mu$ unphysical, the global limit produces a sensible result: 
\begin{equation}
J^\mu = \partial_\nu\, (F^{\mu\nu})\, \theta = J^\mu_{EM} \theta
\label{jackso17}
\end{equation}
In the Weyl case setting the parameter $\theta$ in \eqref{jackso2} to a constant produces a global symmetry. Yet the current still vanishes.

Evidently the vanishing of the Weyl symmetry current both local and {\it a forteriori} also global reflects the particularly peculiar role of the Weyl ``symmetry" in the examined models. We assert that Weyl invariance has no dynamical role in conformal cosmological models based on action \eqref{jackso1} and \eqref{jackso1-2}. At best, a possible calculational convenience may be achieved. 

It is common practice in recent cosmology papers to view the tensor-scalar Lagrangians like \eqref{jackso1} and  \eqref{jackso1-2} as gauge theories presented in the so-called ``Jordan" frame.  Gauge fixing brings them to ``Einstein" frame, with one less minimally coupled scalar field and no Weyl symmetry due to gauge fixing.  In view of the vanishing Weyl current, a better description is that the Einstein-Hilbert theory does not arise  from gauge fixing but from a redefinition of dynamical variables by a spurion field. In detail this works as follows.

We shall now call the metric tensor in Jordan frame \eqref{jackso1} and \eqref{jackso1-2} $ g^J_{\alpha\beta}$  and call the Einstein metric tensor $g^E_{\alpha\beta}$  . Upon substituting in $\mathcal{L}_1, g^J_{\alpha\beta}\, \varphi^2$ by $g^E_{\alpha\beta}$, the spurion field $\varphi$ disappears and the Einstein-Hilbert action, which clearly lacks local conformal symmetry, emerges from \eqref{jackso1} ($G$ has been set to $3/4\pi$):
\begin{equation}
\mathcal{L}^{EH}_1 = \sqrt{-g_E}\, \left\{\frac{1}{2}\ R (g_E)\right\}
\label{jackso18}
\end{equation}
Similarly, substituting $g^J_{\alpha\beta} (\varphi^2 -\psi^2) = g^E_{\alpha\beta}$ in $\mathcal{L}_2$ and parameterizing the scalar fields as $\varphi = u \cosh \omega$ and $\psi = u \sinh \omega$ we obtain (setting $G=3/4\pi$)
\begin{equation}
\mathcal{L}^{EH}_2 = \sqrt{-g_E}\, \left\{\frac{1}{2}\ R (g_E) -\frac{1}{2} \,  g^{\alpha\beta}_E\, \partial_\alpha\, \omega\, \partial_\beta\, \omega \right\}
\label{jackso19}
\end{equation}
where $\omega$ is a physical massless scalar field. The spurion field $u$ has disappeared and so have Weyl and SO(1,1) invariances. SO(1,1) symmetry becomes the shift symmetry $\omega \to \omega + \epsilon$. Correspondingly the SO(1,1) current \eqref{jackso15} reads
\begin{equation}
J^\mu_{\text{SHIFT}} = \epsilon \sqrt{-g_J} g^{\mu\nu}_J\, u^2 \partial_\nu \omega = \epsilon \sqrt{-g_E}\, g^{\mu\nu}_E\, \partial_\nu \omega
\label{jackso20}
\end{equation}

 Introducing a spurion field and dressing up a model to appear gauge invariant is what we call a fake gauge invariance. We note that others have expressed similar or related criticisms \cite{Quiros:2014wda}.
 
 Up to now the only merit of Weyl symmetry in model \eqref{jackso1} is to provide a derivation for the new improved energy-momentum tensor $\theta^{CCJ}_{\alpha\beta}$: varying $I_1$ with respect to $g^{\alpha\beta}$ produces $\theta^{CCJ}_{\alpha\beta}$ in the flat limit \cite{Cal1970}. It will be interesting to find the symmetry current in a conventional Weyl invariant model, built on the square of the Weyl tensor. There the symmetry is again local, but no scalar field is present to absorb the ``gauge freedom."

%

\subsection*{\centerline{Acknowledgement}}
 We thank S. Deser, M. Hertzberg, T. Iadecola and P. Steinhardt for useful discussions. This research was supported in part by U.S. Department of Energy, Grant No. DE-SC0010025 (S.-Y.P.).
 \vskip 1ex
\noindent {\bf Note added.} We have been informed that the symmetry current in the Weyl tensor theory has been  recently examined and it also vanishes \cite{MCap140}.

\end{document}